\pdfoutput=1
\documentclass[pr,twocolumn]{revtex4}
\usepackage{amssymb}
\usepackage{amsmath}
\usepackage{mathptmx}
\usepackage{color}
\usepackage{euscript}
\usepackage{eufrak}
\usepackage{hyperref}
\usepackage{graphicx}
\usepackage{bm}

\def\q{\quad}

\begin{document}

\title{In the diffraction shadow: Norton waves versus surface plasmon-polaritons in the optical region}

\author{ A.~Yu.~Nikitin$^{1,2}$}
\author{ Sergio G. Rodrigo$^1$}
\author{ F.~J.~Garc\'{i}a-Vidal$^3$}
\author{ L.~Mart\'{i}n-Moreno$^1$}
\email{lmm@unizar.es}
\affiliation{$^1$ Instituto de Ciencia de Materiales de Arag\'{o}n and Departamento de F\'{i}sica de la Materia Condensada,
CSIC-Universidad de Zaragoza, E-50009, Zaragoza, Spain \\
$^2$ A. Ya. Usikov Institute for Radiophysics and Electronics, Ukrainian Academy of Sciences, 61085 Kharkov, Ukraine\\
$^3$ Departamento de F\'{i}sica Te\'{o}rica de la Materia Condensada, Universidad Aut\'{o}noma de Madrid, E-28049
Madrid, Spain}

\date{\today}

\begin{abstract}
Surface electromagnetic modes supported by metal
surfaces have a great potential for
uses in miniaturised detectors and optical circuits.
For many applications these modes are excited locally. In the optical regime,
Surface Plasmon Polaritons (SPPs) have been thought to
dominate the fields at the surface, beyond a transition region comprising 3-4
wavelengths from the source. In this work we demonstrate
that at sufficiently long distances SPPs are not the main
contribution to the field. Instead, for all metals,
a different type of wave prevails, which we term Norton waves for their reminiscence to those found in the radio-wave regime at the
surface of the Earth.
Our results show that Norton Waves are stronger at the surface
than SPPs at distances larger than 6-9
SPP's absorption lengths, the precise value depending on wavelength and metal.
Moreover, Norton waves decay more slowly than SPPs in the direction normal to the surface.
\end{abstract}

\maketitle

The confinement of the electromagnetic field associated to Surface Plasmon Polaritons (SPPs), and their
intrinsic speed, make them very interesting candidates for their use in
photonics\cite{Ebbesen04,Maier_book}.
Due to this, the study of the electromagnetic (EM) fields radiated by localized
sources (like defects\cite{Krenn}, nano-gratings or apertures\cite{Genet}) placed on a surface has
received a renewed interest in the last decade. This is an old problem,
which was put at the forefront in the early 1900's by its possible relevance to the transmission of
radio signals.
The seminal works of Zenneck\cite{Zenneck07} and Sommerfeld\cite{Sommerfeld09} unveiled the
existence of surface waves running along the Earth, which can be considered as a lossy dielectric.
The interest in these works fainted after the realization that radio transmission does not occur via the exponentially damped surface modes, but through reflection at the ionosphere. Nevertheless, Norton
subsequently showed that, in the long distance limit, radio waves decay algebraically at the surface\cite{Norton36}.
This result triggered a debate on the range of validity of Zenneck-Sommerfeld and Norton waves in the radio regime that has propagated to our days (see Ref.\cite{Collin04} for more details and an historical account).
Recently, advances in nanofabrication have allowed the scaling down of old radio devices into the
optical regime\cite{Ebbesen08}.
Metallic surfaces are specially interesting because they support SPPs,
which are surface EM modes strongly confined to the plane.
The analysis of the surface EM fields created by a
localized source in a metal surface has revealed the existence of a near-field region,
extending for 3-4 wavelengths, where the field presents a complex
dependence\cite{Tejeira05,LalanneNature06}. SPPs have been thought to dominate the
EM field beyond this region.
In this work we show that,
irrespectively of the metal considered,
the long-distance asymptotic limit of the EM field at metal surface is not the SPP but a
different type of wave, which we denote as Norton waves (NWs) due to their
reminiscence to those
found in dielectric surfaces. We show the range of validity of SPPs
and NWs and the distance and field amplitude after which the latter dominate.

\begin{figure*}[th!]
\includegraphics[width=15cm]{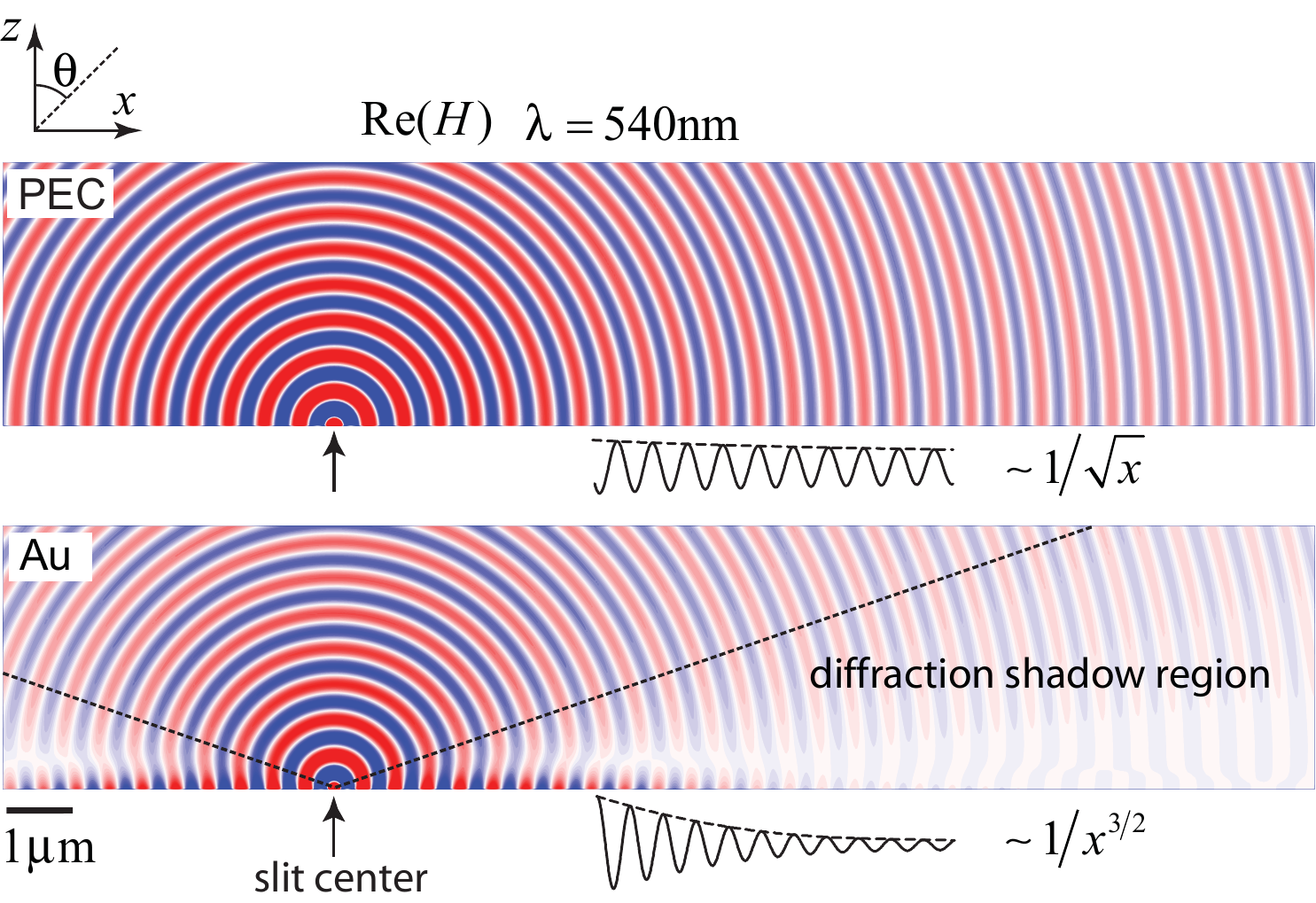}
\caption{Snapshot of the magnetic field radiated by a back-illuminated subwavelength slit
in an optically thick metal film. In the top panel the metal is treated as a Perfect Electrical
Conductor, while the metal in the lower panel is Au. The wavelength is $\lambda=540$nm.}\label{Fig1}
\end{figure*}

%\begin{figure}
%\includegraphics[width=8cm]{Fig1}
%\caption{Snapshot of the magnetic field radiated by a back-illuminated subwavelength slit
%in an optically thick metal film. In the top panel the metal is treated as a Perfect Electrical
%Conductor, while the metal in the lower panel is Au. The wavelength is $\lambda=540$nm.}\label{Fig1}
%\end{figure}

Although we will show later how the obtained results apply to dipole sources, let us concentrate
first on the EM fields
emerging from a subwavelength slit, placed in an optically thick
metal film. The film is back-illuminated by normal-incident p-polarized light
with wavelength $\lambda$ (i.e. the wavevector in vacuum is $g=2 \pi/\lambda$).
The frequency-dependent dielectric
constant of the metal is $\epsilon_m$. This system has been chosen
for analytical simplicity (the full EM field can be expressed in
terms of the magnetic field along the slit axes,
$H_y(X,Z)$) and, also because it is a configuration that has been
amply studied both theoretically\cite{Lalanne05,YoungExpPRL05,Gay06,Leveque07,Besbes07,ShengOptExpr08} and
experimentally\cite{Tejeira_louncherNature07,LalannePRL07,WoggonAPB08,Laluet08}.
Figure 1 renders a snapshot of the radiated $H_y(X,Z)$ (computed with the FDTD method)
for slit width of $A=100$nm and $\lambda=540nm$, for both
a Perfect Electrical Conductor (PEC, characterized  by
$|\epsilon_m|=\infty$) and Au\cite{SergioPRB08}. The choice of metal and wavelength is motivated for
proof-of-principle purposes on the existence of NWs, but we will show later on that our results are
applicable to other metals and frequency ranges.
Our treatment fully takes into account the vectorial nature of the EM fields and, therefore, goes beyond
the scalar approximations considered in other works\cite{Gay06,WoggonAPB08}.

It is apparent from Fig. 1 that the
effect of a finite $\epsilon_m$  is to strongly modify the radiation
pattern close to the metal surface. Although we will provide
expressions for the field everywhere, our main focus will be to
characterize the fields within the diffraction shadow which, loosely
speaking is the region where radiation from a slit in a real metal is strongly
reduced with respect to the PEC case.

The Green's dyadic method is more suitable for an analytical study of this problem.
Within this
method, the field radiated at the point ${\bf R}=(X,Z)$ by a slit of width A is
(see Supplementary Information S1 for the justification of this expression and its
validation with numerical calculations)
\begin{equation}\label{Hslit}
H(x,z) \approx \sqrt{\varepsilon_m-1} \int_{-a/2}^{a/2}\, G(x-x',z)\, E_x(x',z=-\delta)\, dx'
\end{equation}
where the "Green's function" $G(x,z)$ is the magnetic field generated
by a dipolar source with the electric field pointing along the
x-direction, placed at the metal interface, and $\delta$ is the skin depth for the metal.
In this expression and throughout the paper
all distances denoted by lower case letters are expressed in dimensionless units
as $x=gX$,  $z=gZ$ and $a=gA$.
Alternatively, given that the fundamental
waveguide mode inside the slit is constant in the x-direction, $G(x,z)$ can be seen
as the magnetic field radiated by an infinitesimally thin slit. The
angular spectrum representation of this function is:
\begin{equation}\label{Gexacta}
G(x,z) = \int_{-\infty}^\infty D(q) \, e^{iqx+iq_z z} \, dq,
\end{equation}
where $q$ is the x-component of the wavevector (in units of $g$),
$D(q) \, = \,q_{zm}/\left[2\pi \, (\epsilon_m q_z+q_{zm})\right] $, $q_z = \sqrt{1-q^2}$
and $q_{zm} = \sqrt{\epsilon_m-q^2}$ \cite{sign}.

The solution to this integral is not known in closed form. Fortunately, there are
mathematical methods\cite{FelsenMarcuvitz} for extracting its long-distance asymptotic expression, $G_{asymp}(x,z)$. The rigorous calculation for $G_{asymp}(x,z)$ is provided in
the Supplementary Material S2 and, additionally, a simplified derivation will be given later on.
But, before going into the mathematical details, let us now concentrate on the fields
at the metal surface and give the result obtained:
\begin{equation}\label{asymptotic}
G_{asymp}(x,0) = G_{SPP}(x,0) + G_{NW}(x,0).
\end{equation}
In this expression $G_{SPP}(x,0)$ is the SPP contribution
\begin{equation}\label{Gspp}
G_{SPP}(x,0) =  2\pi i\, C_p \, e^{iq_p x},
\end{equation}
where $q_p=\sqrt{\varepsilon_m/(1+\varepsilon_m)}$ is the SPP momentum,
and $C_p=\, q_p^3/\left[ 2 \pi (\varepsilon_m-1)\right] $ is the residue of $D(q)$ at $q_p$.

The second term is
\begin{equation}\label{GNW}
G_{NW}(x,0)= \frac{e^{i x +i\pi/4}}{\sqrt{2 \pi}} \, \frac{\varepsilon_m}{\sqrt{\varepsilon_m-1}}
\,x^{-3/2}.
\end{equation}
As will be shown later, this term is the 2D optical analog in metal
surfaces of the Norton wave\cite{Norton36}
found in the study of the radio-wave radiation of point dipoles on lossy dielectric interfaces.
Dimensionality accounts for the difference between the decay
laws: $x^{-3/2}$ (2D dipoles) and the $x^{-2}$ (3D dipoles).

\begin{figure}[h!]
\includegraphics[width=8cm]{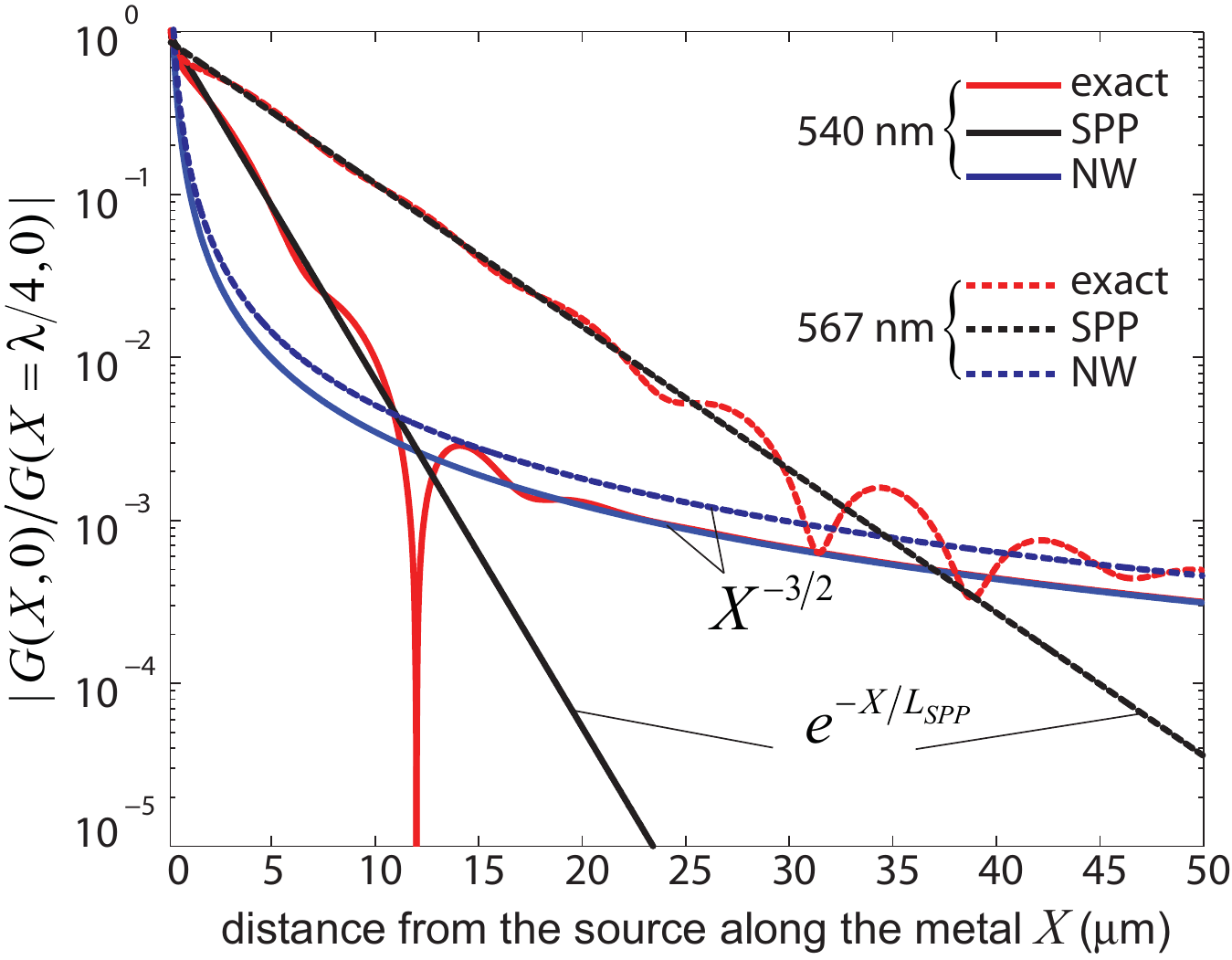}
\caption{The magnetic field at the metal surface radiated by an horizontal dipole
as function of distance.
The dependencies are presented for Au at two wavelengths: 540 nm (continuous curves) and
567 nm (discontinuous curves). The figure shows the exact result (red) and the SPP (black) and
NW (blue) contributions.}\label{Fig2}
\end{figure}

The validity of Eq.~\eqref{asymptotic} and the competition between SPPs and NWs
is illustrated in Fig.~\ref{Fig2}, which shows
the magnetic field at the surface radiated by an infinitesimally thin subwavelength slit, for Au at
two different wavelengths. In each case, this figure
renders the exact result (computed numerically from Eq.~\eqref{Hslit}) and the
the SPP and NW contributions. For the cases considered in this figure, the asymptotic
result given by Eq.~\eqref{asymptotic} is virtually indistinguishable from the exact result even for
$X\approx 3\mu m$ and it is not represented. The field is
mainly SPP-like at the shorter distances, while NW dominates at sufficiently long distances
from the source. Notice that the relative phase of the NW and the SPP contributions at the
distance where their modulus are equal changes with wavelength. So, their destructive interference
may lead to the cancellation of the field (as in Au, at $\lambda=540$nm at $X\approx 12\mu m$) or,
if the cancellation is not
complete, to the appearance of small oscillations
in the total field amplitude of the field (as in the presented case of Au at
$\lambda=567$nm). It is worth noticing that similar oscillations were found in
Scanning Near Field Optical Microscope experiments in Au\cite{WoggonAPB08}, but their
origen was unknown.

\begin{figure}[h!]
\includegraphics[width=8cm]{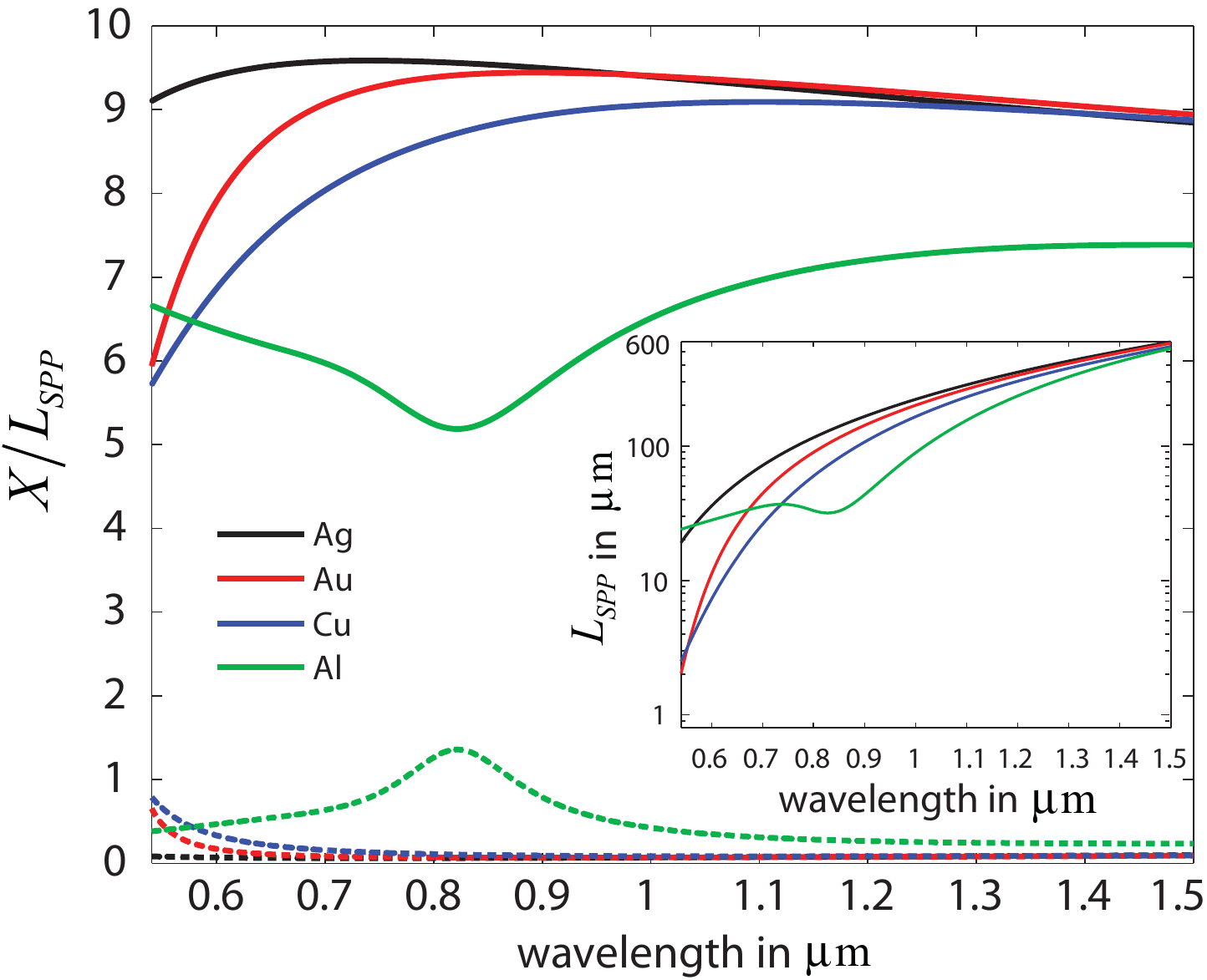}
\caption{Spectral dependence of the crossover between SPP and NW at the metal surface, for different metals.
The continuous lines render $x_{NW}$, defined as the distance at which the amplitude of the NW
is larger than the SPP one. The discontinuous lines render $x_{a}(0.1)$ the minimum distance at which
$G_{asymp}$ gives a relative error of 10\% with respect to the exact result.
The inset renders the spectral dependence SPP absorption length for the considered metals.}
\label{Fig3_xcrossover}
\end{figure}

Beyond the particular examples presented in Fig.~\ref{Fig2}, the expression given
by Eq.~\eqref{asymptotic} is a good approximation for the field at the surface, at sufficiently long distances from the source. In order to quantify this statement, we define $x_{a}(\beta)$ as
the minimum distance such that $|(G(x_a,0)-G_{asymp}(x_a,0))/G(x_a,0)|<\beta$. The discontinuous
lines in Fig.~\ref{Fig3_xcrossover} show the spectral dependence in the optical and telecom regimes of $x_{a}(0.1)$ for different metals, in units of the corresponding SPP absorption length.
Given that the NW decays algebraically and the SPP exponentially with distance, at sufficiently large distances the NW is the main contribution to the field at the surface,
for all metals and all wavelengths. The crossover from NW to SPP is represented
in Figure.~\ref{Fig3_xcrossover}, which renders the spectral dependence
of the distance at which the NW contribution is larger than the SPP one, $x_{NW}$, for different metals.
This distance strongly depends upon the dielectric permittivity of the metal, being smaller for very lossy metals, as Cu and Au in
the region of inter-band transition (close to $\lambda=500$ nm for both metals).

Undoubtedly, the existence of NWs in metal surfaces has passed
unnoticed up to now due to their small amplitude. In order to characterize how
much has the field decayed when the NW takes over, we consider the
ratio $|G(X_{NW},0)/G(X=\lambda/4,0)|$ (the distance $X=\lambda/4$ has been arbitrarily
chosen to give a representative reference in the near-field). In the optical regime, when the NW takes over
the field has decayed by a factor ranging from $10^{-2}$ for lossy metals (like Ni) to $10^{-3}-10^{-4}$ for
Ag. Therefore, the NW is not a good channel for sending information along the surface. Nevertheless,
and given that estimations of the field at the surface far away from the source
based on the decay of SPPs may be orders of magnitude wrong,
NWs may have to be taken into account for precise analysis or design of experiments.

In order to show the origin of NWs, and its relation to other waves discussed in the literature,
as creeping waves and SPPs, let us concentrate on the physical interpretation
of the field radiated by the slit.
Additionally, this will lead to a ``poor man's'' (yet correct) derivation
of some of the main results. It is clear from Eq.~\eqref{Gexacta} that a slit excites the whole
range of diffraction modes (both radiative and evanescent) with an amplitude given by
$D(q)$, which can loosely speaking be understood as the density of EM
modes with a given wavevector $q$ at the slit position\cite{note}. The standard
treatment of $G(x,z)$
in the far-field relies on the observation that, although all modes
are always present, their contributions cancel out due to destructive
interference whenever the phase $\Phi={\bf q r}$ changes rapidly.
Thus, only the region in $q$-space where the phase presents an
extremum contributes to the far field. For a given point $(x,z)$ (or $(r, \theta$) in
polar coordinates with $\theta$ defined as the angle from the normal to the surface),
the extremum occurs at the condition
$(q/q_z)_{min}=x/z$, i.e. $q_{min}=\mathrm{sin}\theta$.
Expanding the integrand around this extremum
leads to the ``ray-optics'' (RO) contribution

\begin{equation}\label{GRO}
\begin{split}
G_{RO}(r,\theta) = \sqrt{\frac{2 \pi}{ r}} \, e^{ir-i\pi/4}\, \cos \theta \,
D(\sin \theta) =\\ \frac{e^{ir-i\pi/4}}{\sqrt{2\pi r}}\, \frac{\cos\theta\sqrt{\epsilon_m-\sin^2\theta}}{\epsilon_m\cos\theta + \sqrt{\epsilon_m-\sin^2\theta}}.
\end{split}
\end{equation}

This analytical result reproduces what was observed in Fig. 1: the magnetic
field radiated by an infinitesimally thin slit in a PEC is isotropic, but the pattern in a real metal
is strongly modified close to the surface, for angles
such that $\mathrm{cos}\theta \lesssim 1/\sqrt{|\epsilon_m|}$.

However, right at the surface the derivative of the phase $\Phi=q x$ never cancels and the
saddle point approximation outlined above can not be directly applied. As the integral of the product of a smooth and rapidly oscillating function is very small, only the parts of the
angular spectrum where $D(q)$ changes rapidly in the scale of $2\pi/x$ will give a net contribution
to the integral. For very small $x$ all the ``density of states'' contribute. As $x$ increases,
the smooth long-$q$ region of $D(q)$
is progressively canceled out in the integral, which is eventually dominated by the strong (and rapid) contribution from the pole in $D(q)$.
The contribution of this pole gives the SPP field.
Notice that, in a lossy metal, the density of states associated to the plasmon pole has a finite width,
which causes the exponential decrease of the SPP amplitude with distance
(characterized by the SPP propagation
length $l_{SPP}=\mathrm{Im}(q_p)^{-1})$.

The previous argument explains why the field at the surface is not the SPP
for all distances and is expected to have a complex dependence with $x$. Recently, Lalanne and coworkers have termed ``creeping wave'' (CW) to the difference between the
exact field and the approximation given by the SPP pole\cite{LalanneNature06}.
The numerical study of the CW has shown that
it is a damped wave which, along the surface, oscillates
with the free-space wavevector and decays after a few wavelengths. A point to notice is that despite the
$e^{i x} $ dependence, the CW arises from the whole angular spectrum, not only from regions close to $q=1$.

\begin{figure}[t!]
\includegraphics[width=8cm]{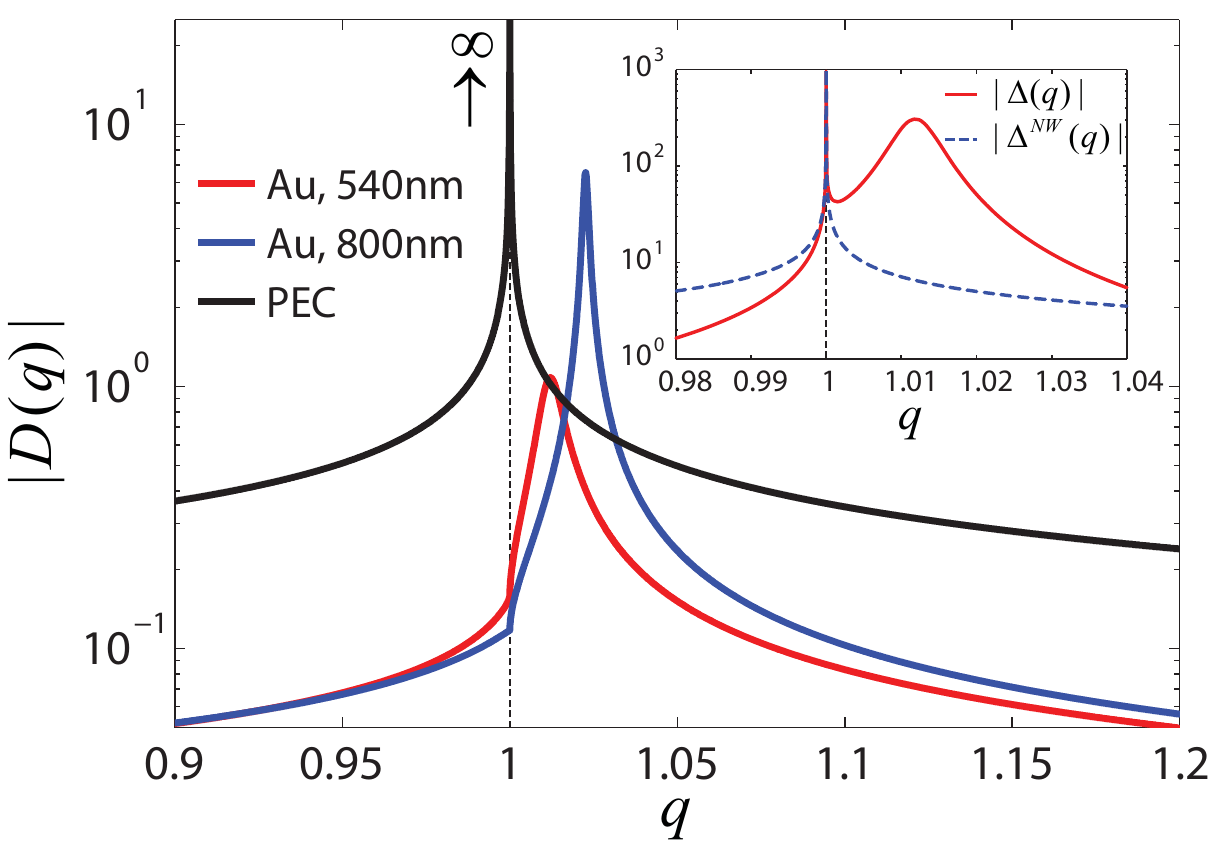}
\caption{The modulus of
$D(q)$, defined in Eq.~\eqref{Gexacta}. The blue and red curves are for the Au surface
(at $\lambda=800$nm and $\lambda=540$nm, respectively), whereas the
green curve is for the PEC ($\xi=0$).} The inset shows the same for the function
$\Delta(q)$ defined in the text.\label{Fig4_DOS}
\end{figure}

However, the SPP pole is not the sharpest feature of $D(q)$: the derivative of $D(q)$ diverges at the branch
point $q_z=0$. This is
illustrated in Fig.~\ref{Fig4_DOS}, which shows that $|D(q)|$ has a kink at $q=1$. The contribution to the integral from this kink is expected to be small but, as the kink can not be characterized
by a typical width in q-space, it is not as strongly suppressed as the SPP contribution
when integrated with an oscillatory function.
In order to show that the kink originates the NW, it is convenient to integrate by parts $G(x,0)$. Then, from Eq.\eqref{Gexacta} we obtain
$G(x,0) = (i / x) \, \int_{-\infty}^\infty \Delta(q) \, e^{iqx} \, dq$, with $\Delta(q)= q \, G'(q)$.
This representation has the advantage that the kink in $D(q)$  transforms into a square root
singularity (see inset to Fig.~\ref{Fig4_DOS}). The contribution close to $q_z=0$ can be retrieved by keeping
the singularity but setting $q_z=0$ everywhere else,
this is, by defining $ \Delta^{NW}(q)=( \varepsilon_m/\left[ 2 \pi \sqrt{\varepsilon_m-1}\right] )\, (1/q_z)$.
The inset to Fig.~\ref{Fig4_DOS} renders the comparison between $\Delta(q)$ and $ \Delta^{NW}(q)$,
for a representative case.
Of course, $ \Delta^{NW}(q)$ is only a good approximation to $ \Delta(q)$ close to $q=1$, so its use for integration over the whole angular spectrum could seem unjustified. However for very large $x$ this is valid,
as only the region close to $q=1$ contributes.
With this, $G_{NW}(x,0) = (i/x)\, \int_{-\infty}^\infty  \,\Delta^{NW}(q)\, e^{iqx}
\, dq \, = i (\epsilon_m / \left[ 2 \sqrt{\varepsilon_m-1}\right] ) H_0^{(1)}(x)/x$.
Recalling that this expression is only valid for large $x$, we
substitute $H_0^{(1)}(x)$ for its asymptotic value and obtain the result in Eq.~\eqref{GNW}.
The motivation for our terminology on this type of wave is that the asymptotic term
found by Norton, for the case of radio waves emitted by a dipole in a dielectric\cite{Norton36},
also decays algebraically and
originates from the angular spectrum close to $q=1$.
The SPP contribution can also be extracted
from the previous representation by expanding $\Delta_q$ close to $q=q_p$. As the NW and the SPP arise from different parts of the angular spectrum, their fields can
be directly added up, leading to Eq.~\eqref{asymptotic}.

The relevance of NWs with respect to SPPs increases when we move away from the surface. Since
for $z=0$ the NW originates
from $q$-values close to the light-line, its decay with distance to the surface is expected to be slower than the exponential decay of SPPs. In order to obtain the dependence of the Norton wave on both $x$ and $z$, we have carried out the asymptotic analysis of the Green's function given by Eq.~\eqref{Hslit}, using the general method described in Ref.\cite{FelsenMarcuvitz}. The derivation and full asymptotic form can be found in the Supplementary Material S2. The result up to terms of the order $r^{-1/2}$ was already presented in Ref.\cite{ShengOptExpr08}, where it was shown that the long-distance asymptotic expressions are excellent approximations even for distances as small as $x=1$ (i.e., $X=\lambda/(2\pi)$). However, the asymptotic expression given by Ref.\cite{ShengOptExpr08} misses some contributions of the order $r^{-3/2}$, so it can not be used for the problem discussed in this paper. We find that the expression for $G(x,z)$ in the far-field ($r\gg1$) is (see Supplementary Material S2):
\begin{equation}\label{Gxz}
G_{asymp}(x,z) = G_{SPP}(x,z) + G_{RO}(x,z) + G_{NW}(x,z),
\end{equation}
where $G_{SPP}(x,z) = G_{SPP}(x,0)\, e^{i q_{pz} z}$, with $q_{pz}=-1/\sqrt{(1+\varepsilon_m)}$,
$G_{RO}(x,z)$ is given by Eq.~\eqref{GRO},
and $G_{NW}(x,z)$, defined as the term that goes as $r^{-3/2}$, is given by
\begin{equation}\label{GNWxz}
G_{NW}(x,z)  = \frac{e^{ir+ i \pi/4}}{\sqrt{2 \pi} }\,
\left\lbrace \frac{d^2}{d\phi^2} \left[\frac{- D(\sin\phi)
\cos(\phi)}{2 \cos((\phi-\theta)/2)}\right]\right\rbrace_{\phi=\theta} \,
r^{-3/2}.
\end{equation}

The difference between the exact and asymptotic
expressions $\Delta G(x,z)=G(x,z)-G_{asymp}(x,z)$ decays as $r^{-5/2}$.

\begin{figure}[h!]
\includegraphics[width=8cm]{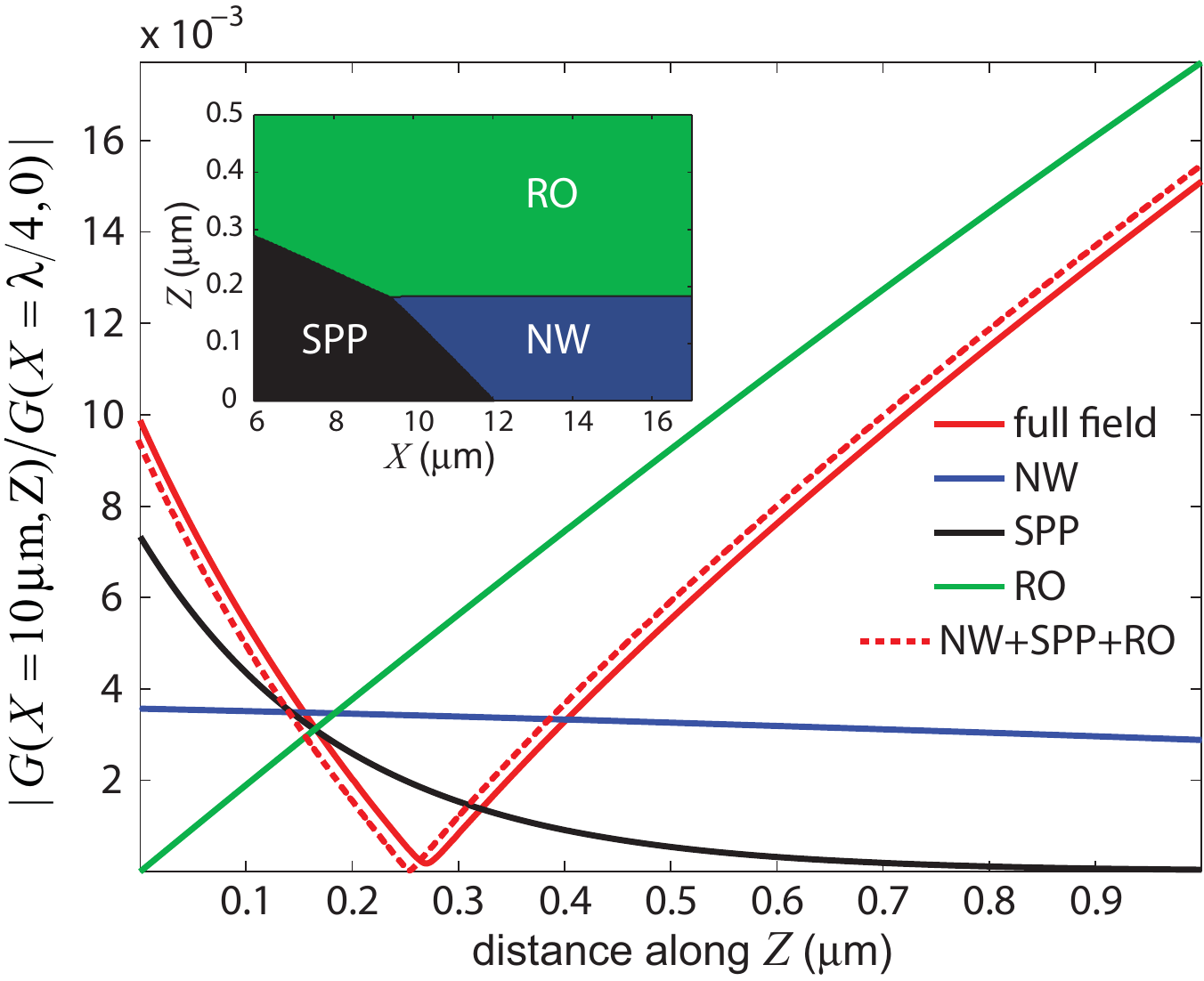}
\caption{The magnetic field close to the metal surface for Au at  $\lambda=540$ nm.
The main figure shows the dependence with $z$ for fixed $X=10\mu m$. The exact result $G(X,Z)$
(red curve) and the asymptotic one, $G_{asymp}(X,Z)$ (red dashed),
are presented, together with the contribution from
the SPP (black curve), ray optics (green curve) and NW (blue line).
The inset shows
the regions in the $X-Z$ plane where the different terms dominate.}
\label{Fig5}
\end{figure}

Figure.~\ref{Fig5} presents the comparison between the $z$-dependence of the
exact $G(x,z)$ and $G_{asymp}(x,z)$, at $X=10 \mu m$.
These results show that the asymptotic expression is very accurate.
Also that, as expected, the NW contribution decays with distance to the
surface much more slowly than the SPP one.
In both chosen examples the SPP dominates right at the surface, but the NW takes over at a finite distance from it. However, at sufficiently large $z$ the RO contribution always dominates. The inset to Fig.~\ref{Fig5} shows,
for Au at $\lambda=540$nm,  which of the three ``asymptotic'' terms (SPP, RO and NW) dominates in the $X-Z$ plane. This inset shows that the NW is the largest contribution over a ``stripe'' close to the surface. As the region where the NW is dominant satisfies $z \ll x$, the full expression for the NW given by Eq.~\eqref{GNWxz} can be approximated by (for $|\varepsilon_m|\gg1$)
\begin{equation}\label{NW}
\begin{split}
|G_{NW}(x,z)| \, \simeq  \, |G_{NW}(x,0)| \, \frac{1}{|1+\sqrt{\varepsilon_m} z/x|^3}.
\end{split}
\end{equation}
The comparison of this expression with that of the RO contribution allows for an estimation of the distance to the surface at which the
crossover between NW and RO occurs, $z_{NW}$. We obtain
$z_{NW}=|\sqrt{\varepsilon_m}|/|1+\sqrt{\varepsilon_m} z/x|^3$ which,
if $z_{NW}/x \ll 1/|\sqrt{\varepsilon_m}|$, implies $z_{NW}\approx|\sqrt{\varepsilon_m}|$, or $Z_{NW}\approx\lambda |\sqrt{\varepsilon_m}|/(2\pi)$. Notice that the algebraic decay of the NW with $z$, reflects that this wave arises from the interference of its constituent components.

To summarize, we have shown that, in the asymptotic limit of long distances to the source, the SPPs are not the main channel for EM fields at the metal surface.
Instead, after a few SPP absorption lengths, Norton waves take over. This occurs for any metal
and any frequency range. Norton waves decay much
more slowly than SPPs both along the surface (as $x^{-3/2}$ for 2D dipoles) and along the perpendicular direction.

\begin{acknowledgments}
The authors acknowledge support from the Spanish Ministry of Science under projects MAT2008-06609-C02 and CSD2007-046-Nanolight.es.
\end{acknowledgments}

\appendix

\section{Field in the vacuum half-space using the Green's dyadic}

Consider a p-polarized electromagnetic wave (with wavelength $\lambda$ and wavevector $g = 2\pi/\lambda$)
incident
onto a metal film with a subwavelength slit.
The metal film is optically thick and extends
from $z=-W$ (where the EM field impinges) to $z=0$ (the exit side). The dielectric constant of the metal is
$\varepsilon_m$.
The slit has width $A$ and we set the origin of the $x$-axis at the center of the slit.

According to Lippmann-Shwinger integral equation\cite{Martin95}, the electric field
at any point at exit side of the film ($z>0$) is given by the following integral relation
\begin{equation}\label{sa1}
\mathbf{E}(\mathbf{R}) = \mathbf{E}_0(\mathbf{R}) +g^2 \int_Vd\mathbf{R}'\, \Delta\varepsilon(\mathbf{R}') \, \hat{G}^E(\mathbf{R},\mathbf{R}') \, \mathbf{E}(\mathbf{R}'),
\end{equation}
where $\mathbf{E}_0(\mathbf{r})$ is the solution without the slit,
$\Delta\varepsilon(\mathbf{r})=1-\varepsilon_m$ in the volume occupied by the slit, $V$, and zero everywhere else.

In the case of an optically thick film the field $\mathbf{E}_0(\mathbf{R})$ can be neglected at the exit side
and the dyadic $\hat{G}^E(\mathbf{R},\mathbf{R}')$ can be approximated by the one corresponding to a single metal-vacuum interface.
In order to obtain the magnetic field from Eq.~\eqref{sa1},
we use the Maxwell equation $\mathbf{H}=(- i/g)\nabla_\mathbf{R} \times \mathbf{E}$
and arrive at
\begin{equation}\label{sa2}
\mathbf{H}(\mathbf{r}) = \int_Vd\mathbf{r}' \, \Delta\varepsilon(\mathbf{r}') \, \hat{G}^H(\mathbf{r},\mathbf{r}') \, \mathbf{E}(\mathbf{r}'),
\end{equation}
where we have passed to dimensionless distances $\mathbf{r} = g \mathbf{R}$. The dyadic $\hat{G}^H$ connects the magnetic field outside of the slit with the electric field inside the slit. For the considered two-dimensional geometry, where only p-polarized waves are involved, the magnetic field $\mathbf{H}$ points along the $y$-direction.
Assuming that the electric field inside
the slit mainly points along the $x$-direction only the $yz$ element
of the dyadic $\hat{G}^H$ needs be computed.
We denote this element by $\hat{G}^H_{yx}(x,z;x',z')$.
Following Ref.~\onlinecite{Novotny}, we find that
\begin{equation}\label{sa3}
\hat{G}^H_{yx}(x,z;x',z')= \frac{i}{2\pi}\int dq\frac{q_{zm}}{\varepsilon_mq_z+q_{zm}} e^{iq(x-x')+iq_zz-iq_{zm}z'}
\end{equation}
where $q$ is the dimensionless $x$-component of the wavevector, $q=k_x/g$,  $q_z = \sqrt{1-q^2}$
and $q_z = \sqrt{\epsilon_m-q^2}$.

The integrant contains the exponential factor $e^{-iq_{zm}z'}$, which decays in the distance of a skin depth
$\delta=1/\mathrm{Im}(q_{zm})$, which is of the order of a few tens of nm in the optical regime. Therefore, the
integration limits in $z'$ can be extended to $[-\infty,0]$. Moreover,
the variation of the Green's dyadic is much faster than that of the electric field
inside the slit, so the electric field
inside the slit can be approximated by its value at the distance $z=-\delta$ (this is obtained as the average distance to the surface, weighted by the exponential decay of the field). An additional advantage of using the field at a short distance inside the slit is that the numerical problems related to the treatment of
corners are eliminated.

The integration over $z'$ can be performed in the following way
\begin{equation}\label{sa4}
\int_{\mathrm{slit}}e^{-iq_{zm}z'}dz'\simeq \int_{-\infty}^0e^{-iq_{zm}z'}dz'=\frac{i}{q_{zm}}
\end{equation}.

We then obtain
\begin{equation}\label{sa4_2}
H_y(\mathbf{r}) =  (\varepsilon_m-1) \int_{-a/2}^{a/2}dx'G_{slit}(x-x',z)E_{x'}(x',z=-\delta),
\end{equation}
with
\begin{equation}\label{sa4_3}
G_{slit}(x,z)=\frac{1}{2\pi}\int dq\frac{1}{\varepsilon_mq_z+q_{zm}} e^{iqx+iq_zz}.
\end{equation}

Since we are interested at the field close to the surface and away from the source (which, as we will show,
arises from the region in the angular spectrum close to $q=1$) and for optical frequencies (where $\varepsilon_m$ is large), this result can be related to the field radiated by a dipolar source $G(x,y)$
at the metal surface:
\begin{equation}\label{sa5}
\begin{split}
H_y(x,z) \simeq \sqrt{\varepsilon_m-1}\int_{-a/2}^{a/2}dx'G(x-x',z)E_{x'}(x',z=-\delta),
\end{split}
\end{equation}

where

\begin{equation}\label{sa5_1}
G(x,z)=\frac{1}{2\pi}\int dq\frac{q_{zm}}{\varepsilon_mq_z+q_{zm}} e^{iqx+iq_zz}.
\end{equation}

The relation between Eq.~\eqref{sa4_2} and Eq.~\eqref{sa5}  can be easily seen by noticing that
$q_{zm}\approx \sqrt{\varepsilon_m-1} $ in the region of interest.

While Eq.~\eqref{sa5} involves an additional approximation, we have preferred to work with the
Green's dyadic for a dipole source (rather than with $G_{slit}$) due to its wider applicability to other problems. In any case, the differences for the slit case are minimal and the methods described in this paper
could be straightforwardly applied to $G_{slit}$.

\begin{figure}[h!]
\includegraphics[width=8cm]{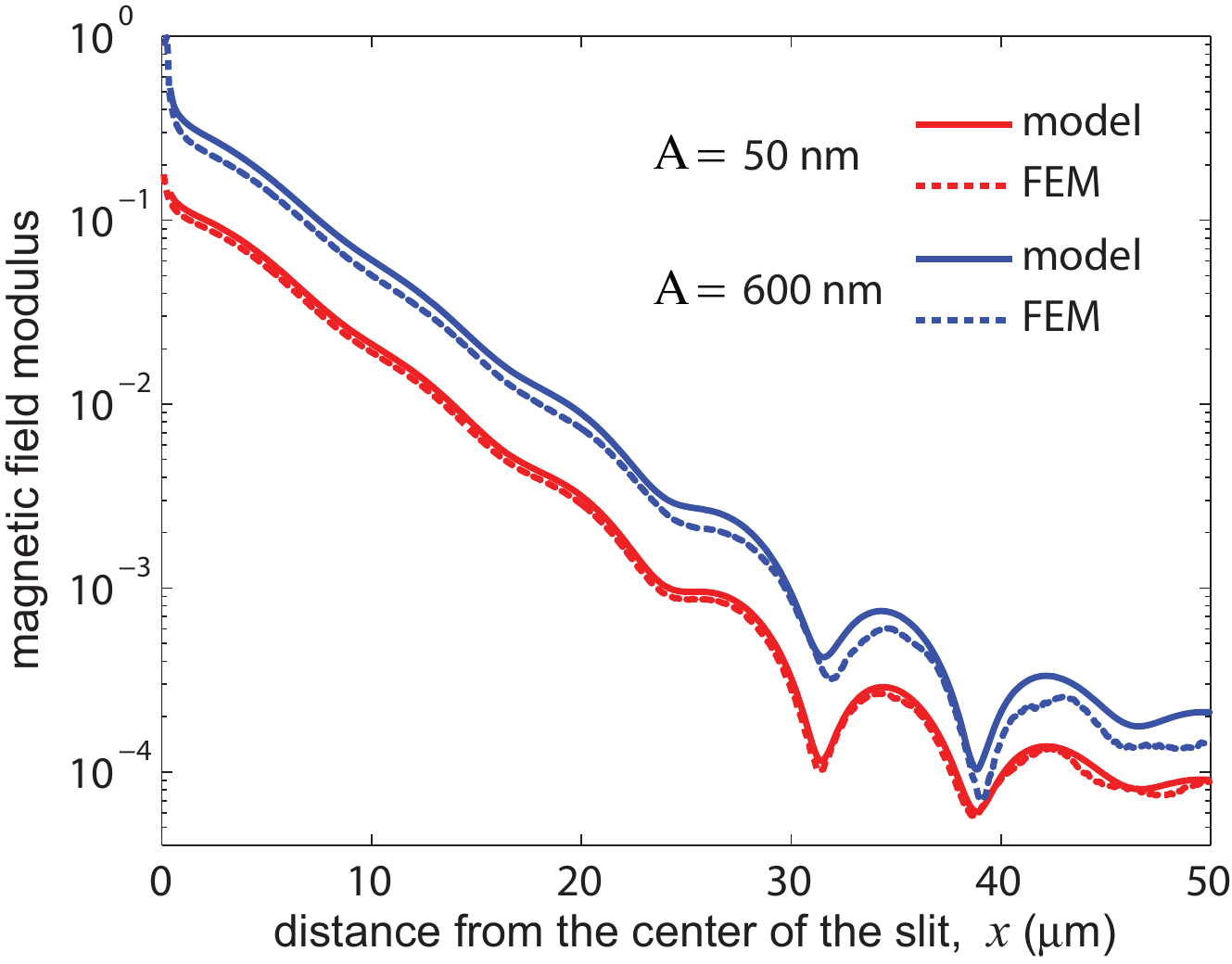}
\caption{Comparison between the FEM and the analytic computation for the magnetic field on the gold-vacuum interface emerging from a slit. The slit widths considered are $50$ nm and $600$ nm, and the wavelength is $567$ nm. The dashed lines are for the FEM calculations, while the continuous lines correspond to the approximate result using Eq.~\eqref{sa5}.}
\label{FigS1}
\end{figure}

In order to validate the expression \eqref{sa5}, we first performed full-vectorial computations using the Finite Element Method (FEM) of the magnetic field emerging from slits with the thicknesses $50$ nm and $600$ nm. The wavelength was chosen to be $576$ nm and the metal of the film is gold. Then we computed the integral given by Eq.~\eqref{sa5} extracting the electric field inside the slit at $z=-\delta$ from the FEM calculations. The function $G(x,z)$ was substituted by its asymptotic value, see the next section S2. The comparison is shown in Fig.~\ref{FigS1}.

These results clearly show that Eq.~\eqref{sa5} is very accurate for subwavelegth slits ($A<<\lambda$) and
even provide a good approximation for $A\sim \lambda$.

\section{Asymptotic behavior of the field: steepest descent method}

Consider the Green's function for a dipole placed at the metal surface given by Eq.~\eqref{sa5_1}.
In this section we sketch the asymptotic analysis performed, which has been done
following the general method described in \cite{FelsenMarcuvitz} for treating Sommerfeld integrals.

In this method, the integrant is first prolonged into the complex $q$-plane. Subsequently,
the following changes of variable are performed $q = \sin\phi$ and $s = \sqrt{2}e^{i\frac{\pi}{4}}\sin\left(\frac{\phi-\theta}{2}\right)$. In polar coordinates ($x=r\sin\theta$
and  $y=r\cos\theta$) the integral takes the following form
\begin{equation}\label{s2}
G = e^{ir}\int_{C}ds \Phi(s)e^{-rs^2}
\end{equation}
with
\begin{equation}\label{s3}
\Phi(s)= \frac{1}{2\pi} \frac{\sqrt{2}e^{-i\pi/4}}{\cos[\frac{\phi(s)-\theta}{2}]}\cdot\frac{\cos[\phi(s)]\sqrt{\epsilon_m-\sin^2[\phi(s)]}}{\epsilon_m\cos[\phi(s)] + \sqrt{\epsilon_m-\sin^2[\phi(s)]}},
\end{equation}
where the contour $C$ in the $s$-plane corresponds to the real axis in the initial plane $q$.
\begin{figure}[h!]
\includegraphics[width=8cm]{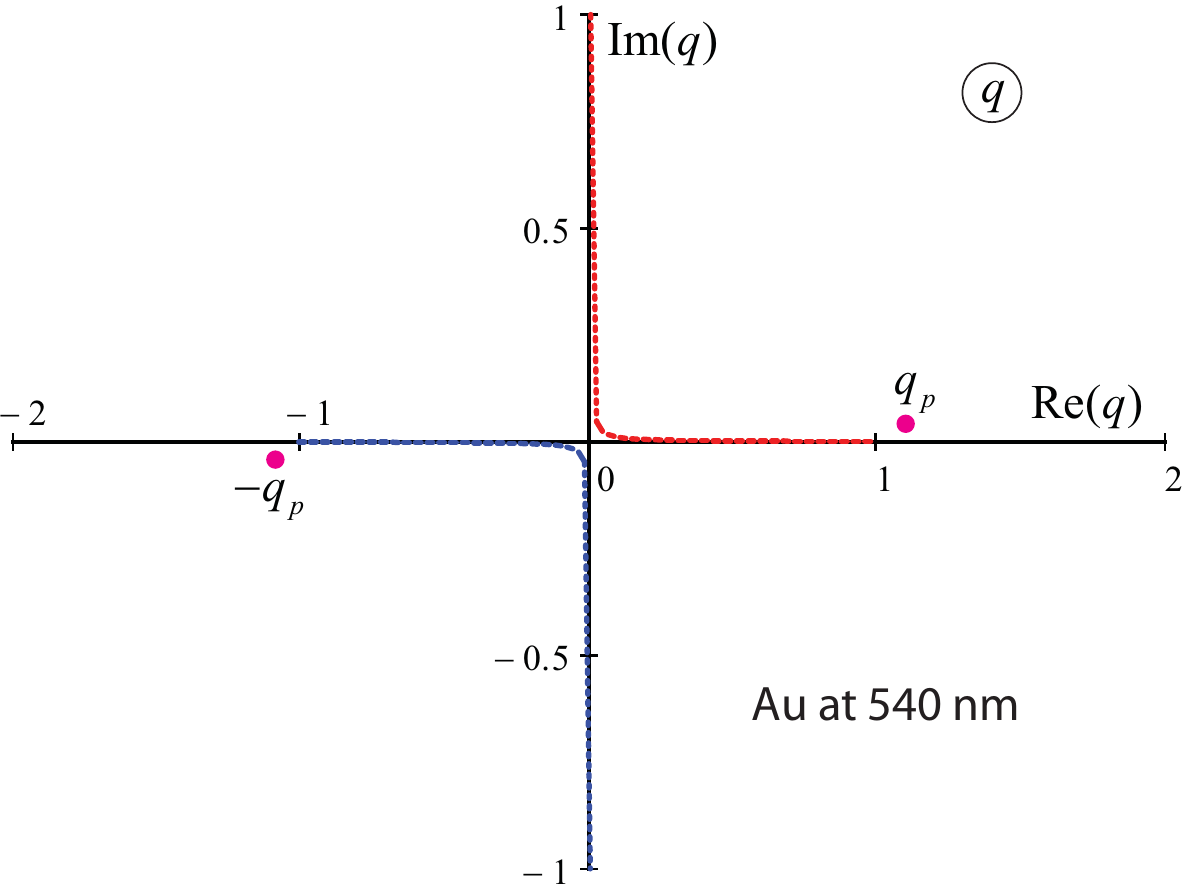}
\caption{The poles $q= \pm q_p$ and the branch cuts $\mathrm{Im}(q_z)=0$ with infinitesimally small absorption of the vacuum in the complex plane $q$. The poles correspond to the dots. For $x>0$ only the pole $q=q_p$ contributes to the result.}
\label{FigS2}
\end{figure}

\begin{figure}[h!]
\includegraphics[width=8cm]{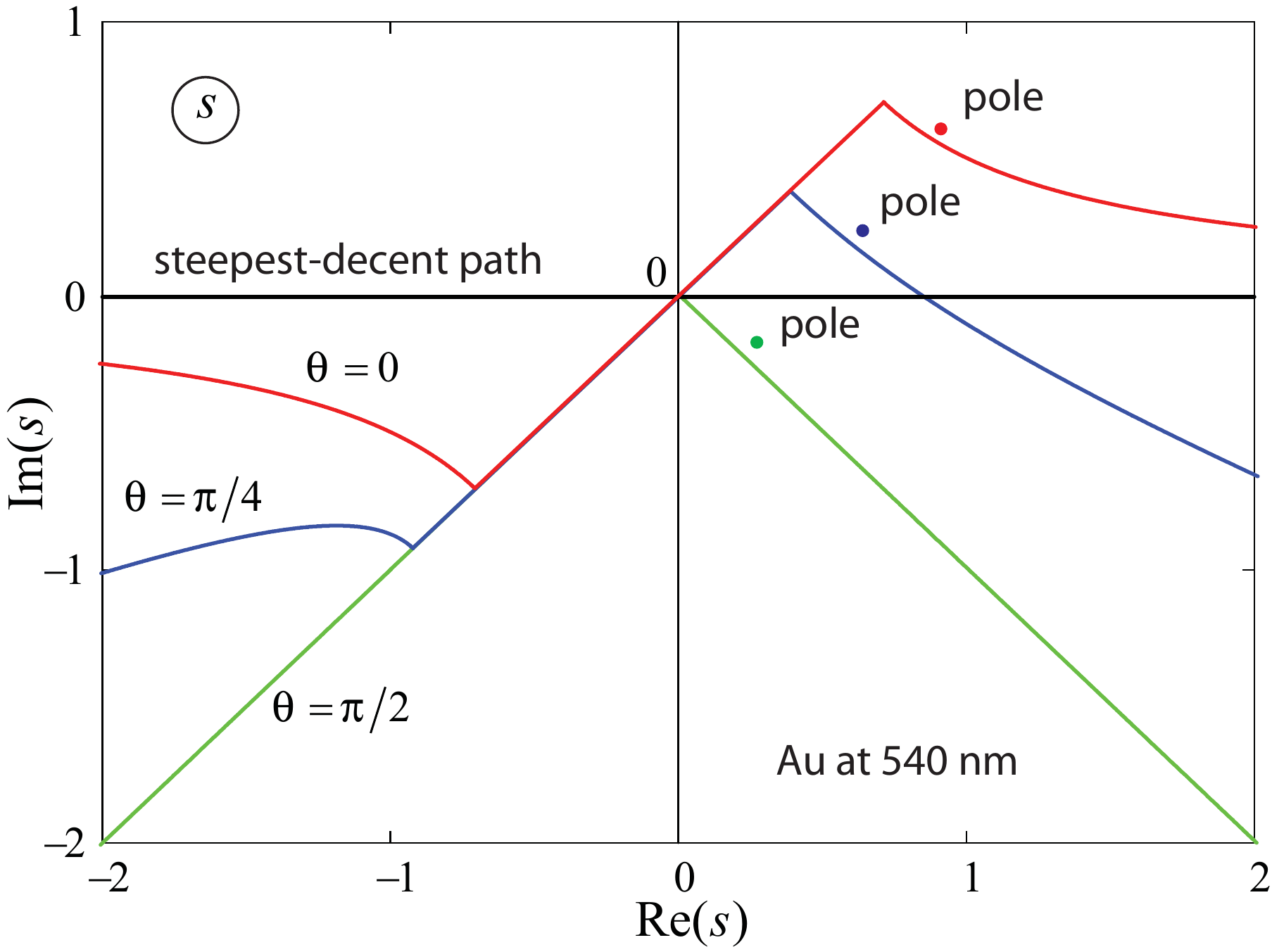}
\caption{The contours of the integration in the complex plane $s$ for different angles $\theta$.
The positions of the poles $s_p$ are mark the dots. When the pole is crossed during the transformation of the initial path to the steepest-decent one, the residue must be taken into account.}
\label{FigS3}
\end{figure}

Then the integrant is separated into singular and non-singular parts
\begin{equation}\label{s4}
\Phi(s) = \frac{C_p}{s-s_p} + \Phi_0(s)
\end{equation}
with
\begin{equation}\label{s5}
\Phi_0(s) = \frac{\Phi(s)(s-s_p) -C_p}{s-s_p},
\end{equation}
where $C_p$ is the residue given by $C_p = \frac{\epsilon\sqrt{\epsilon}}{2\pi(\epsilon^2-1)\sqrt{1+\epsilon}}$, and the position of the pole in the complex plane $s$ is $s_p = \sqrt{2} e^{\imath \pi/4} \mathrm{sin}(\frac{\phi_p-\theta}{2})$,
where $\phi_p = \arccos\left(\frac{-1}{\sqrt{\epsilon+1}}\right)$. Then, after deforming the integral path to the steepest descent one (real axis in the plane $s$), the singular part yields the complementary error function $i\pi C_p e^{r(i - s_p^2)}\mathrm{erfc}(- is_p\sqrt{r})$ with the argument being the  square root of the ``numerical distance'' introduced by Sommerfeld. The non-singular part of the integral can be expanded in Taylor series close to the saddle point $s=0$ providing the infinite sum of the integrals of the Gaussian type. The final result reads
\begin{equation}\label{s6}
G =  i\pi C_p e^{i\mathbf{r}\mathbf{q}_p}\mathrm{erfc}(- is_p\sqrt{r}) +   e^{ir}\sum\limits_{n\in\mathrm{even}}\frac{\Gamma(\frac{n+1}{2})}{n!r^\frac{n+1}{2}}\frac{d^n\Phi_0}{ds^n}|_{s=0}.
\end{equation}
The function $\Phi_0$ is composed of two parts: $\Phi$ and $C_p/(s_p-s)$. Note that the part of the sum in \eqref{s6} coming from the term $C_p/(s_p-s)$ coincides, up to a sign, with the asymptotic expansion of the complementary error function for large arguments without the first term (while the last
term coincides with the residue contribution into the integral). This expansion reads
\begin{equation}\label{s6.1}
\mathrm{erfc}(- is_p\sqrt{r}) =   2\Theta(\theta-\theta_p) + i\frac{e^{rs_p^2}}{\pi}\sum\limits_{n\in\mathrm{even}}\frac{\Gamma(\frac{n+1}{2})}{(s_p^2r)^\frac{n+1}{2}},
\end{equation}
where $\Theta$ is a Heaviside step function, and $\theta_p$ is the angle defining the diffraction shadow, $\theta_p = \mathrm{Re}(\phi_p) - \arccos\left(1/\cosh[\mathrm{Im}(\phi_p)]\right)$. This is the critical angle such that for $\theta>\theta_p$ we have $\mathrm{Im}(s_p)<0$ and the initial transformation of the integration path into the steepest-decent one leads to the crossing of the pole, so that the residue must be taken into account, see Fig.~\ref{FigS2}. An example of the critical angle limiting the diffraction shadow is represented in Fig. 1 of the article by dashed lines.

Retaining the residue contribution, $G_{SPP}$, and the first two terms $G_{RO}$ (proportional to $r^{-1/2}$) and $G_{NW}$ (proportional to $r^{-3/2}$) in the sum coming from $\Phi$ we obtain an expression that is correct in the far-field, up to terms $\Delta G=O(r^{-5/2})$. The result can be rewritten as
\begin{equation}\label{s7}
\begin{split}
G =  G_{SPP} + G_{RO} + G_{NW} + \Delta G.
\end{split}
\end{equation}
The plasmonic term is
\begin{equation}\label{s8}
\begin{split}
G_{SPP}  = 2\pi i C_pe^{i\mathbf{r}\mathbf{q}_p},
\end{split}
\end{equation}
the ``ray optics'' term  is
\begin{equation}\label{s9}
\begin{split}
G_{RO}  = \frac{e^{ir-i\pi/4}}{\sqrt{2\pi r}}\cdot \frac{\cos\theta\sqrt{\epsilon_m-\sin^2\theta}}{\epsilon_m\cos\theta + \sqrt{\epsilon_m-\sin^2\theta}},
\end{split}
\end{equation}
the Norton wave is given by
\begin{equation}\label{s10}
\begin{split}
G_{NW} = \frac{1}{2}\frac{e^{ir-i\frac{3\pi}{4}}}{r\sqrt{2\pi r}} \cdot \frac{d^2}{d\phi^2}\left[\frac{1}{\cos(\frac{\phi-\theta}{2})}\frac{\cos\phi\sqrt{\epsilon-\sin^2\phi}}{\epsilon\cos\phi + \sqrt{\epsilon-\sin^2\phi}}\right]_{\phi=\theta},
\end{split}
\end{equation}
and the rest is
\begin{equation}\label{s11}
\begin{split}
\Delta G = i\pi C_p e^{r(i - s_p^2)}[\mathrm{erfc}(- is_p\sqrt{r})-2] + \\ e^{ir}\sum\limits_{n\in\mathrm{even}}\frac{\Gamma(\frac{1+n}{2})}{n!r^\frac{1+n}{2}}\left(\sigma_n\frac{d^{n}\Phi(s)}{ds^{n}}- \frac{d^n}{ds^n}\frac{C_p}{s-s_p}\right)_{s=0}, \\
\sigma_n =
\left\{
\begin{array}{c}
1, \q n \geq 4 ,\\
0, \q n < 4 .
\end{array}
\right.
\end{split}
\end{equation}


\begin{thebibliography}{99}
\bibitem{sign} The sign in the square root must be taken
so that $Im(q_z)\geq 0$, in order to satisfy the radiation condition.
\bibitem{note} Actually the density
of EM modes with wavevector $q$ at the surface is related to $Im(D(q))$.
\end{thebibliography}
\end{document}